\documentclass[12pt]{article}
\usepackage{amsmath,amssymb,amsfonts,epsfig,graphicx}

\newcommand{\bse}{\begin{subequations}}
\newcommand{\ese}{\end{subequations}}
\newcommand{\be}{\begin{equation}}
\newcommand{\ee}{\end{equation}}
\newcommand{\bea}{\begin{eqnarray}}
\newcommand{\eea}{\end{eqnarray}}
\newcommand{\ba}{\begin{array}}
\newcommand{\ea}{\end{array}}

\makeatletter \@addtoreset{equation}{section}

\makeatother
\def\by{\times}
\def\l{\left}
\def\r{\right}
\def\bl{\biggl}
\def\br{\biggr}

\def\A0{{\cal A}_0}
\def\H{{\bf H}}
\def\N{{\cal N}}

\def\G{{\cal G}}
\def\J{{\mathbb J}}
\def\L5{{\cal L}_5}
\def\M{{\cal M}}

\def\Q{{\cal Q}}
\def\R{{\cal R}}
\def\Z{{\cal Z}}

\def\Tr{{\rm Tr\ }}
\def\super{$PSU(2|2)\times PSU(2|2)\times U(1)$}
\def\ie{{\it i.e.\ }}
\begin{document}
\baselineskip 18pt%

\begin{titlepage}
\vspace*{1mm}%
\hfill%
\vbox{
    \halign{#\hfil        \cr
           IPM/P-2005/080 \cr
           hep-th/0512037 \cr
           } 
      }  
\vspace*{15mm}%

\centerline{{\Large {\bf Extensions of $AdS_5\times S^5$ and the
Plane-wave Superalgebras}}} \centerline{{\large{\bf and}}}
\centerline{{\Large {\bf Their Realization in the Tiny Graviton Matrix Theory}}}%
\vspace*{5mm}
\begin{center}
{\bf M. Ali-Akbari$^{1,2}$, M. M. Sheikh-Jabbari$^{1}$, M. Torabian$^{1,2}$}%
\vspace*{0.4cm}

{\it {$^1$Institute for Studies in Theoretical Physics and Mathematics (IPM)\\
P.O.Box 19395-5531, Tehran, IRAN\\
$^2$Department of Physics, Sharif University of Technology\\
P.O.Box 11365-9161, Tehran, IRAN}}\\
{E-mails: {\tt aliakbari, jabbari, mahdi@theory.ipm.ac.ir}}%
\vspace*{1.5cm}
\end{center}

\begin{center}{\bf Abstract}\end{center}
\begin{quote}
In this paper we consider all consistent extensions of the
$AdS_5\times S^5$ superalgebra, $psu(2,2|4)$, to incorporate brane
charges by introducing both bosonic and fermionic (non)central
extensions. We study the In\"{o}n\"{u}-Wigner contraction of the
extended $psu(2,2|4)$ under the Penrose limit to obtain the most
general consistent extension of  the plane-wave superalgebra and
compare these extensions with the possible BPS (flat or spherical)
brane configurations in the plane-wave background. We give an
explicit realization of some of these extensions in terms of the
Tiny Graviton Matrix Theory (TGMT)\cite{TGMT} which is the $0+1$
dimensional gauge theory conjectured to describe the DLCQ of strings
on the $AdS_5\times S^5$ and/or the plane-wave background.
\end{quote}%
\end{titlepage}
\section{Introduction}%
Historically supersymmetry (SUSY) algebras arose as extensions of
Poincare algebra of a D-dimensional space-time by fermionic
generators, the supercharges. The first extension to the
supersymmetry algebra constructed in this way, the super-Poincare
algebra, appeared when the number of supercharges were taken to be a
multiple  (usually taken to be a power of two) of  dimension of the
smallest spinor representation of the corresponding Poincare
algebra. Such extensions usually come under the title of
$\N$-extended SUSY algebras \cite{HLS}. In this extended versions of
superPoincare the bosonic part of the superalgebra is extended by an
``internal'' $R$-symmetry group, which in four dimensions is a
$U(\N)$ symmetry, under  which the supercharges are in fundamental
representations. It is then natural to ask for other possible
extensions of the $\N$-extended algebras. If such an extension
exists, then we end up with an extended SUSY algebra which contains
the original algebra as a subalgebra. Generically speaking these
extensions can appear in the bosonic or fermionic sectors of the
SUSY algebra and they can be central or non-central. Central
extensions are those which are at the center of the corresponding
Poincare algebra and hence they are necessarily scalars in
space-time. There are also $p$-form extensions, which are all
non-central (for example see \cite{Townsend} and references
therein).  These $p$-form extensions, however, generically commute
with the momenta and the super(Poincare)-charges and hence central
in this sense.\footnote{As discussed e.g. in \cite{Townsend}, an
extension of the SUSY algebra which is not central, upon dimensional
reduction, may appear as a central extension in the lower
dimensional superalgebra.} As is well-known presence of these
$p$-form extensions are necessitated by dualities of string
theories, as they correspond to the charges of extended objects,
$p$-branes, e.g. see \cite{Townsend, Sezgin, Bars}.

For the \textit{non-central} extensions of a given superalgebra one
should check the closure  and the Jacobi identities for the extended
algebra. Generically, as first noted in \cite{M-algebra}, closing
the algebra amounts to  adding fermionic as well as bosonic
$p$-forms, i.e. fermionic brane charges. Existence of the
super-$p$-form charges are necessary for  construction of
supersymmetric Wess-Zumino terms in $p$-brane actions
\cite{M-algebra}.

It is not always necessary to build a SUSY algebra on a Poincare or
Lorentz algebra and one can construct supersymmetric versions of all
classical Lie algebras by adding fermionic generators in the spinor
representations of these algebras and demanding their closure.
Superalgebras obtained in this way all fall into Kac-Nahm
classification of superalgebras \cite{Nahm, Kac}. These algebras
become of great importance in string theory once we consider
theories on spaces which are not asymptotically flat, but still
supersymmetric.

Regardless of the dimension of space-time in which the superalgebra
is defined the maximal number of supercharges in a physically viable
theory, which does not contain higher than spin two fundamental
particles, cannot exceed 32. In this paper hence our focus would be
on the maximal superalgebras which are not based on superPoincare.
The most famous of examples of such cases are the maximally
supersymmetric $AdS_p\times S^q$ $(p,q)= (5,5),\ (4,7)$ and $(7,4)$
geometries and the corresponding plane-waves in ten or eleven
dimensions \cite{Blau}. For the $AdS_5\times S^5$ the relevant
algebra is $psu(2,2|4)$ which will be reviewed and explicitly
presented in section \ref{section2.1}. As has been discussed in
\cite{Gomis-etal, Japanese-school, Lee} to find the maximally
possible extension of this algebra one should add other fermionic,
as well as $p$-form generators. Although pure $psu(2,2|4)$ is a
subsuperalgebra of $ops(1|32)$, it is not the case for its maximal extension.
While the bosonic part of the maximally extended $PSU(2,2|4)$ is a subgroup of bosonic part of $OSp(1|32)$, the whole superalgebra is a contraction of $osp(1|32)$. This algebra will be
presented and discussed in some detail in \ref{section2.2}.
Interestingly and intriguingly $osp(1|32)$ also appears as the
maximal possible extension of all the other known flat space maximal
superalgerbas, namely ten dimensional $\N=2$ superPoincare
\cite{Sezgin, Bars}, eleven dimensional $\N=1$ superPoincare
\cite{M-algebra, Bergshoeff-VanProeyen}.

In this paper we will study the maximal extensions of the $AdS$
superalgebras and the behavior under the Penrose limit. In this way
we will obtain the most general extension of the other class of
maximally supersymmetric non-flat supergravity backgrounds, namely
the plane-wave geometries. This is done in section 3. In this paper
we will discuss the $psu(2,2|4)$  in detail. The computations for
the $osp(8^*|4)$ and $osp(4|8)$ cases can be performed in a  similar
way. The extended eleven dimensional plane-wave superalgebra has
been discussed in \cite{Peeters}.

The extension of the plane-wave superalgebras are of particular
interest because for these cases we have matrix theory formulations
for the DLCQ of type IIB strings and/or the M-theory on these
backgrounds \cite{TGMT, BMN}. As discussed in \cite{TGMT} both of
these matrix theories are ``tiny graviton matrix theories'', the
tiny three brane and membrane theories, respectively. Within these
matrix models one may then find explicit representations of these
algebras and their super-extensions in terms of matrices. These are
the charges corresponding to BPS objects in these matrix models, and
hence the extended brane-type objects in the corresponding string or
M-theories. In section 4, we will focus on the representation of the
extensions of the ten dimensional plane-wave superalgebra in the
tiny graviton matrix theory (TGMT) and their brane interpretation.

In the last section we conclude by summary of our results, outlook
and remarks. In the two appendices  we have gathered a review of the
relevant fermionic notations employed in the paper and a brief
review on the tiny graviton matrix theory.
\section{Extensions of the $psu(2,2|4)$ superalgebra}\label{section2}%
In this section we review the $psu(2,2|4)$ algebra and its possible
extensions. We employ the fermionic notations developed in
\cite{plane-wave-review}. To be self-contained, we have explained
our fermionic notations in the appendix A.

{\bf\textit{ Note on the classification of superalgebras:}}\\
According to the Nahm classification of superalgebras \cite{Nahm},
generically in any superalgebra, besides the central generators, two
kind of bosonic groups are involved. For example in the four
dimensional $\N$-extended SUSY algebras, the four dimensional
Poincare $ISO(3,1)$ and the $U(\N)$ R-symmetry group. The
supercharges are then in the spinor representation of both of these
groups. (The $U(1)$ part of the $U(\N)$ R-symmetry is generically
anomalous in the supersymmetric gauge theories). This can be
generalized by taking the supercharges to be in spinor
representation of any two classical Lie groups. The most famous
cases are $su(m|n)$ algebras in which the $2mn$ (real) supercharges
are in the spinor representation of $su(m)$ and $su(n)$. That is, we
can denote the supercharges by $Q_{a\alpha}$, where $a,\alpha$,
$a=1,2,\cdots, m,\ \alpha=1,2,\cdots, n$ are the fundamental indices
of $SU(m)$ and $SU(n)$. The bosonic part of this superalgebra,
besides $su(m)\times su(n)$, generically contains an extra $U(1)$.
For the special case of $m=n$, however, this $U(1)$ factor becomes
central to the whole algebra and maybe extracted out. In which case
to emphasize  absence of the $U(1)$, the algebra is denoted by
$psu(m|m)$. The examples of this algebras are $psu(2,2|4)$ whose
bosonic part is $su(2,2)\simeq so(4,2)$ times $su(4)\simeq so(6)$,
and $su(4|2)$ with the bosonic part $su(4)\times su(2)\times
u(1)\simeq so(6)\times so(3)\times u(1)$ \cite{SU(4|2)}. The former
is a superalgebra with $2\times 4\times 4=32$ supercharges and the
latter with $2\times 4\times 2=16$.

The next supergroups/algebras relevant to the 11 dimensional AdS
cases are $OSp(m|n)$. The bosonic part of which are $so(m)$ (or more
precisely $Spin(m)$) times $USp(n)$. (In our conventions
$USp(2n)\simeq Spin(2n+1)$.)  The examples of these algebras
appearing in the M-theory backgrounds are $osp(8|4)$ with the
bosonic part $so(8)\times usp(4)\simeq so(8)\times so(3,2)$,
$osp(8^*|4)$ whose bosonic part is $so(8^*)\times usp(4)\simeq
so(6,2)\times so(5)$ and finally $osp(1|32)$ with the bosonic part
$usp(32)$. All these superalgebras have 32 supercharges. Further
discussions on these superalgebras may be found in \cite{spinor}.

\subsection{Unextended $psu(2,2|4)$ superalgebra}\label{section2.1}%
$PSU(2,2|4)$ is the supergroup of the $\N=4$ four dimensional
Yang-Mills superconformal field theory. As explained above, in the
most natural notations,  the supercharges of this algebra carry two
six dimensional Weyl indices of $so(6)$ and  $so(4,2)$. That is, the
supercharges are labeled as $Q_{I\hat J}$, where $I$ and ${\hat
J}=1,2,3,4$ are respectively the Weyl indices of $so(6)$ and
$so(4,2)$. This is the notations and conventions used in
\cite{plane-wave-review}.\footnote{Note, however that, although the
most convenient one, this is not the notation usually used for
labeling the supercharges of this algebra. Usually the $Q_{I\hat J}$
supercharges are decomposed in terms of $so(3,1)$ spinors, and as
superPoincare and super conformal charges $S$, see e.g.
\cite{Freedman}. The latter labeling seems more suitable from the
$\N=4$ super-Yang-Mills theory.} The superalgebra in this notation
takes the compact form of \cite{plane-wave-review}%
\bse\label{psu(2,2|4)}\begin{align}
[\J_{AB},\J_{CD}]=i(\delta_{A[C}\J_{B]D}-\delta_{B[C}\J_{A]D})\  , \
\ \ [\J_{\hat\mu\hat\nu},\J_{\hat\rho\hat\lambda}]&=i(
\eta_{\hat\mu[\hat\rho}\J_{\hat\nu]\hat\lambda}-\eta_{\hat\nu[\hat\rho}
\J_{\hat\mu]\hat\lambda})\\ \ \ \cr \{\Q_{I\hat{J}},
\Q^{\dagger{K\hat{L}}}\}=2(i\gamma^{\hat\mu\hat\nu})_{\hat I}^{\
\hat K}{\J}_{\hat\mu\hat\nu}\ \delta_{J}^{\ {L}} +
2(i\gamma^{AB})_{J}^{\ L}{\J}_{AB}\ &\delta_{\hat I}^{\ \hat K}\ ,\
\ \ \{\Q_{I\hat{J}}, \Q_{K\hat{L}}\}=0
\\ \ \ \cr
[\J_{AB}, \Q_{I\hat{J}}]=(i\gamma_{AB})_{I}^{\ K} \Q_{K\hat{J}}\ \ \
\ , \ \ \ [\J_{\hat\mu\hat\nu},&
\Q_{I\hat{J}}]=(i\gamma_{\hat\mu\hat\nu})_{\hat J}^{\ \hat L}
\Q_{I\hat{L}} \end{align}\ese%
where $A,B=1,2,\cdots, 6$ are the fundamental $SO(6)$ indices and
$\J_{AB}$ are generators of $so(6)$. The hatted Greek indices run
over $-1,0,\cdots,4$ and are fundamental $SO(4,2)$ indices, whose
algebra generators are denoted by $\J_{\hat\mu\hat\nu}$ and
$i\gamma_{AB}$, $i\gamma_{\hat\mu\hat\nu}$ are the commutators of
the $SO(6)$ and $SO(4,2)$ $\gamma$-matrices in the Weyl
representation and hence they are $4\times 4$ hermitian matrices.
For more details see Appendix A.

\subsection{The extended $psu(2,2|4)$ superalgebra}\label{section2.2}%
In this subsection we study the most general superalgebra permitted
by just group theory considerations, we add additional objects
(terms) which fall in different allowed representations of the
bosonic part of the superalgebra. Later we give physical
interpretations for this extra terms using field theory description,
and also realizations in the langauge of the corresponding matrix
theory. The physical significance of these extensions lies in the
fact that they correspond to charges of different extended objects.
Explicitly, in order to have supersymmetric (BPS) objects in the
theory, we have to add their charges into the superalgebra.

According to our conventions, as discussed in Appendix
\ref{Notations}, spinorial supercharges carry two indices
corresponding to fundamentals of $su(2,2), su(4)$ as $\Q_{I\hat J}$
which is the direct product of two representations ${\bf
4}\otimes{\bf \hat 4}$. When we anticommute $\Q, \Q^\dagger$, each
fundamental properly multiplies with its antifundamental counterpart
to give%
\[ {\bf 4}\otimes{\bf\bar 4} = {\bf 1}\oplus{\bf 15} \]%
Therefore, group theory dictates that the allowed extensions which
can appear in the right-hand-side of the $\Q,\Q^\dagger$
anticommutator should fall into the following representations of
$SO(4,2)\times S(6)$:
\be ({\bf 1}\oplus{\bf 15}) \otimes ({\bf 1}\oplus{\bf 15}) = ({\bf
1}, {\bf 1})\oplus ({\bf 1},{\bf 15})\oplus({\bf 15}, {\bf
1})\oplus({\bf 15}, {\bf 15})
\ee%
Similarly the extensions to  $\Q,\Q$ anticommutator, as
\[
{\bf 4}\otimes{\bf 4} = {\bf 6}_a\oplus{\bf 10}_s,
\]%
must fall into \be {\rm Sym}[({\bf 6}_a\oplus{\bf 10}_s) \otimes
({\bf 6}_a\oplus{\bf 10}_s)] = ({\bf 6}_a,{\bf 6}_a)
\oplus ({\bf 10}_s,{\bf 10}_s) \ee%
of $SO(4,2)\times S(6)$. Note that $({\bf 6}_a,{\bf 10}_s)\oplus
({\bf 10}_s,{\bf 6}_a)$ being antisymmetric cannot appear in the
$\Q,\Q$ anticommutator.

Therefore, the most general extended anticommutators read as%
\be\label{QQdagger-extended}\begin{split} \{\Q_{I\hat J},\Q^{\dagger
K\hat L} \} &= \chi\ \delta_I^{\ K} \ \delta_{\hat J}^{\ \hat L}\cr
&+ \J_{AB}(i\gamma^{AB})_I^{\ K}\ \delta_{\hat J}^{\ \hat L} +
\delta_I^{\ K} \ \J_{\hat\mu\hat\nu}
(i\gamma^{\hat\mu\hat\nu})_{\hat J}^{\ \hat L} \cr &+
\R_{AB\hat\mu\hat\nu}\ (i\gamma^{AB})_I^{\ K} \
(i\gamma^{\hat\mu\hat\nu})_{\hat J}^{\ \hat L}
\end{split}
\ee \be\label{QQ-extended}\begin{split} \{\Q_{I\hat J},\Q_{K\hat L}
\} &= \Z_{A\hat\mu}\ (\gamma^A)_{IK} \ (\gamma^{\hat\mu})_{\hat
J\hat L} \cr &+ \Z^{++}_{ABC\hat\mu\hat\nu\hat\rho}\
(\gamma^{ABC})_{\{IK\}} \ (\gamma^{\hat\mu\hat\nu\hat\rho})_{\{\hat
J\hat L\}}
\end{split}\ee
Note that $\gamma^A_{IK}$ and $\gamma^{\hat\mu}_{{\hat J}{\hat L}}$
are antisymmetric in $IK$ and ${\hat J}{\hat L}$ indices while
$\gamma^{ABC}$ and $\gamma^{\hat\mu\hat\nu\hat\rho}$ are symmetric.
$\Z^{++}_{ABC\hat\mu\hat\nu\hat\rho}$ is self-dual in both $SO(6)$
and $SO(4,2)$ indices, that is \be\label{6d-selfdual}
\frac{1}{3!}\epsilon^{ABC}_{\ \ \ \ DEF}\
\Z^{++}_{ABC\hat\mu\hat\nu\hat\rho}=+\Z^{++}_{DEF\hat\mu\hat\nu\hat\rho}\
,\ \ \ \frac{1}{3!}\epsilon^{\hat\mu\hat\nu\hat\rho}_{\ \ \ \
\hat\alpha\hat\beta\hat\lambda}\
\Z^{++}_{ABC\hat\mu\hat\nu\hat\rho}=+\Z^{++}_{DEF\hat\alpha\hat\beta\hat\lambda}\
. \ee Since $6_a$ and $10_a$ representations of $SU(4)$ and
$SU(2,2)$ are complex valued the corresponding extensions,
$\Z_{A\hat\mu}$ and $\Z^{++}_{ABC\hat\mu\hat\nu\hat\rho}$ are also
complex while $\chi$ and $\R_{AB\hat\mu\hat\nu}$, as well as
$\J_{\hat\mu\hat\nu}$ and $\J_{AB}$ are hermitian.

The necessary and sufficient condition for the consistency of the
extended algebra, is that it satisfies different Jacobi identities
among fermionic/bosonic  generators. The extensions, except for the
$\chi$, carry $SO(6)$ and $SO(4,2)$ tensor indices and hence have
non-trivial commutators with $\J_{\hat\mu\hat\nu}$ and $ \J_{AB}$
generators \textit{i.e.} they are \textit{not} central to the
original $psu(2,2|4)$ superalgebra. Therefore, closure of the
algebra and the Jacobi identities forces non-zero commutators
between the supercharges $Q$ and the extensions $\Z$ and $\R$.
Moreover, in order to close the algebra we should also add more
fermionic generators, {\it i.e.} fermionic counterparts to the
extensions or the fermionic brane charges. This phenomenon was first
noted by Sezgin in the  context of the M-theory superalgebra
\cite{M-algebra} and then extended to the other cases. For more
details and the structure of the new (anti)commutators see
\cite{Japanese-school, Lee, Bergshoeff-VanProeyen,  Peeters}.

Physically, $\chi, \R_{AB\hat\mu\hat\nu}, \Z_{A\hat\mu},
\Z^{++}_{ABC\hat\mu\hat\nu\hat\rho}$ respectively correspond to
charges of D-instanton, D3, F1/D1, NS5/D5 branes. From the effective
field theory description and its Wess-Zumino term, it can be shown
that they are spatial integral of total derivatives, and are
topological rather than Noether charges \cite{Gomis-etal}. In fact
in the above extensions, the \textit{spatial} components of the
extensions corresponds to brane \textit{dipole} moment charges. For
example, as we will see explicitly in section 4 from  the tiny
graviton Matrix theory construction $\R_{AB\hat\mu\hat\nu}$ leads to
the RR dipole moment of spherical (giant graviton) D3-branes.

One may consider extensions of the eleven dimensional $AdS$
superalgebras, $osp(8|4,\mathbb{R})$ and $osp(8^*|4)$. It has been
argued that the maximal extension of both of these lead to a contraction of
$osp(1|32)$ \cite{Gomis-etal, Bergshoeff-VanProeyen, Peeters}. Intriguingly, a different contraction of the same $osp(1|32)$ algebra appears  as the maximal extension  of $psu(2,2|4)$ we
discussed above. (In our above discussions we did not add the other
needed fermionic generators to make the $osp(1|32)$ structure explicit. More discussions on this
maybe found in \cite{Japanese-school, Lee}.) It seems that there is
a uniqueness theorem \cite{Gomis-etal, Bergshoeff-VanProeyen}, that the maximal
extension of any superalgebra with 32 supercharges is either
$osp(1|32)$ or a contraction of that.

\section{In\"{o}n\"{u}-Wigner contraction of the extended $psu(2,2|4)$} %
Parallel to the Penrose limit which, at the level of the geometry,
takes us from the $AdS_5\times S^5$ solution to the plane-wave
solution, there is a complementary process, known as
In\"{o}n\"{u}-Wigner contraction \cite{IW}, which does the same at
the level of the (super)algebra \cite{Hatsuda}. In this section,
following notations of \cite{plane-wave-review}, we first briefly
review the action of the Penrose limit over the pure (unextended)
super-isometry group of the $AdS_5\times S^5$ which exactly produces
super-isometry group of the plane-wave and then contract the
extended super-isometry group of the $AdS_5\times S^5$ with all the
possible extensions to obtain the most general extensions of the
plane-wave superalgebra. In this way we can trace back the
extensions of the plane-wave algebra to their counterparts in the
$AdS$ and hence read their physical interpretations.

\begin{itemize}\item {\it Contraction of the bosonic part}\end{itemize}%
The bosonic part of isometries is $SO(4,2)\times SO(6)$ with the
generators $\J_{\hat\mu\hat\nu}$ and $\J_{\hat A\hat B}$. In order
to contract the algebra, it is more convenient to
decompose them as%
\bse\begin{align} \J_{\hat\mu\hat\nu} &= \l\{J_{ij},\ L_i =
\frac{1}{R}(J_{-1i}+iJ_{0i}),\ K_i = \frac{-1}{R}(J_{-1i}-iJ_{0i}),\
\mu R^2P^++\frac{1}{2\mu} {\bf H} = J_{-10}\r\} \\
\J_{\hat A\hat B} &= \l\{J_{ab},\ L_a =
\frac{1}{R}(J_{5a}+iJ_{6a}),\ K_a = \frac{-1}{R}(J_{5a}-iJ_{6a}),\
\mu R^2P^+-\frac{1}{2\mu}{\bf H}
= J_{56}\r\},\end{align}\ese%
where $i,a=1,2,3,4$. In the above parametrization the Penrose limit
is $R\rightarrow\infty$ while keeping $J_{ij}, J_{ab}, L_i, K_i,
L_a, K_a, P^+$ and ${\bf H}$ fixed.

\begin{itemize}\item {\it Contraction of the fermionic part}\end{itemize}%
In order to perform the contraction on the fermionic part it is
convenient to adopt the $SO(4)\times SO(4)$ notation for the
fermions and scale them as follows (see \cite{plane-wave-review} or
the Appendix \ref{A2}), \be\label{Q-decompose-rescale}
 \Q_{I\hat J} \rightarrow (\sqrt\mu R\ q_{\alpha\beta}, \sqrt\mu
R\ q_{\dot\alpha\dot\beta}, \frac{1}{\sqrt\mu} Q_{\alpha\dot\beta},
\frac{1}{\sqrt\mu}\  Q_{\dot\alpha\beta}),
\ee%
send $R\to \infty$ while keeping the $q$ and $Q$ in the
right-hand-side fixed.

With the above decomposition and scaling the superalgebra
\eqref{psu(2,2|4)} decomposes into the dynamical and kinematical
superalgebras with the following anticommutators
\be\label{qq-anticommutator} \{q_{\alpha \beta},q^{\dagger\rho
\lambda}\}=P^+\delta_{\alpha}^{\ \rho} \delta_{\beta}^{\ \lambda}\ ,
\ \ \ \{q_{\alpha \beta},q^{\dagger\dot\alpha \dot\beta}\}=0\ ,\ \
\{q_{\dot\alpha \dot\beta},q^{\dagger\dot\rho
\dot\lambda}\}=P^+\delta_{\dot\alpha}^{\
\dot\rho}\delta_{\dot\beta}^{\ \dot\lambda}\  , \ee
\bea\label{qQ-anticommutator} \{q_{\alpha \beta},Q^{\dagger\dot\rho
\lambda}\}=i (\sigma^i)_{\alpha}^{\ \dot\rho} \delta_{\beta}^{\
\lambda} K^i\ &,& \ \ \ \{q_{\alpha \beta},Q^{\dagger\rho
\dot\lambda}\}=i (\sigma^a)_{\beta}^{\ \dot\lambda}
\delta_{\alpha}^{\ \rho} K^a\ , \cr \{q_{\dot\alpha
\dot\beta},Q^{\dagger\dot\rho \lambda}\}=i (\sigma^a)_{\dot\beta}^{\
\lambda} \delta_{\dot\alpha}^{\ \dot\rho} L^a\ &,& \ \ \
\{q_{\dot\alpha \dot\beta},Q^{\dagger\rho \dot\lambda}\}=i
(\sigma^i)_{\dot\alpha}^{\ \rho} \delta_{\dot\beta}^{\ \dot\lambda}
L^i\ , \eea \bea\label{QQ-anticommutator} \{Q_{\alpha
\dot\beta},Q^{\dagger\rho \dot\lambda}\}&=&\ \delta_{\alpha}^{\
\rho} \delta_{\dot\beta}^{\ \dot\lambda}\ {\bf H} + \mu
(i\sigma^{ij})_{\alpha}^{\ \rho} \delta_{\dot\beta}^{\ \dot\lambda}\
J^{ij} + \mu (i\sigma^{ab})_{\dot\beta}^{\
\dot\lambda}\delta_{\alpha}^{\ \rho} J^{ab} \ , \cr
\{Q_{\alpha \dot\beta},Q^{\dagger\dot\rho \lambda}\}&=& 0 \ , \\
\{Q_{\dot\alpha \beta},Q^{\dagger\dot\rho \lambda}\}&=&\
\delta_{\dot\alpha}^{\ \dot\rho} \delta_{\beta}^{\ \lambda}\ {\bf H}
+\mu (i\sigma^{ij})_{\dot\alpha}^{\ \dot\rho} \delta_{\beta}^{\
\lambda}\ J^{ij} + \mu (i\sigma^{ab})_{\beta}^{\
\lambda}\delta_{\dot\alpha}^{\ \dot\rho} J^{ab} \ . \nonumber \eea
As it is seen from \eqref{QQ}, the dynamical supercharges
$Q_{\alpha\dot\beta}, Q_{\dot\alpha\beta}$ form a
subalgebra/subgroup of the original $PSU(2,2|4)$, which can be
identified as $PSU(2|2)\times PSU(2|2)\times U(1)_{\bf H}$. This is
a superalgebra with 16 supercharges and is in fact the
super-isometry of the recently obtained and explored ten dimensional
LLM \cite{LLM} solutions.

After this brief review of the contraction of the unextended
$AdS_5\times S^5$ superalgebra,  we now study the  contraction of
the most extended superalgebra \eqref{QQdagger-extended},
\eqref{QQ-extended}. We follow a similar logic. That is, first we
decompose the fermionic and bosonic generators into the irreducible
representations of $SO(4)\times SO(4)$ and then scale each
representation in the appropriate way. The supercharges $Q$ should,
of course, be decomposed and scaled as in
\eqref{Q-decompose-rescale}. The bosonic form-field extensions
should then be scaled as:%
\bse\label{Penrose-on-extensions}\begin{align}
 \R_{-1i5a}-\R_{0i6a} =
R^{2}c_{ia} &\quad,\quad \R_{-1i5a}+\R_{0i6a} = C_{ia} \\
\R_{-1i6a}+\R_{0i5a} = R^{2}\hat{c}_{ia} &\quad,\quad
\R_{-1i6a}-\R_{0i5a} = \hat{C}_{ia} \\ \Z_{ia}+4\Z_{-10i56a} =
R^2d_{ia} &\quad,\quad \Z_{ia}-4\Z_{-10i56a} = D_{ia} \\
\Z_{-15}-\Z_{06} = R^{2}c &\quad,\quad \Z_{-15}+\Z_{06} = C \\
\Z_{-16}+\Z_{05} = R^{2}\hat{c} &\quad,\quad \Z_{-16}-\Z_{05} =
\hat{C} \\ \chi+\R_{-1056} = R^{2}z &\quad,\quad \chi-\R_{-1056}=Z\
, \end{align}\ese%
while keeping $c$'s and $C$'s fixed when sending $R$ to infinity. We
will discuss the scaling of other $\R_{abij}$ components and various
components of $\Z_{ABC\hat\mu\hat\nu\hat\rho}^{++}$ in the next
subsection.

\subsection{Extensions of the dynamical superalgebra}

The superalgebra \eqref{QQdagger-extended}, \eqref{QQ-extended}
decomposes into dynamical and kinematical parts under the
In\"{o}n\"{u}-Wigner contraction. The dynamical supercharges again
form a subalgebra given by the following (anti)commutation relations

\bse\label{dynamical-extended-1}\begin{align}
\{Q_{{\alpha}\dot{\beta}},Q^{\dagger\rho\dot{\lambda}}\} &=
\delta_{\alpha}^{\rho} \delta_{\dot{\beta}}^{\dot{\lambda}}(\H+Z)
+\mu(i\sigma^{ij})_{\alpha}^{\rho}
\delta_{\dot{\beta}}^{\dot{\lambda}}{\bf J}_{ij}
+\mu(i\sigma^{ab})_{\dot{\beta}}^{\dot{\lambda}}
\delta_{\alpha}^{\rho}{\bf J}_{ab} \cr &+
\mu\delta_{\alpha}^{\rho}(i\sigma^{ab})_{\dot{\beta}}^{\dot{\lambda}}
\R_{ab}-\mu(i\sigma^{ij})_{\alpha}^{\rho}\delta_{\dot{\beta}}^{\dot{\lambda}}\R_{ij}
-\mu(i\sigma^{ij})_{\alpha}^{\rho}(i\sigma^{ab})_{\dot{\beta}}^{\dot{\lambda}}\R_{ijab}^{+-}
\\ \{Q_{\alpha\dot{\beta}},Q_{\rho\dot{\lambda}}\} &=
-\mu\delta_{\alpha\rho}\delta_{\dot{\beta}\dot{\lambda}}(C+i\hat{C})
-4\mu(i\sigma^{ij})_{\alpha\rho}(i\sigma^{ab})_{\dot{\beta}\dot{\lambda}}{\M}^{+-}_{ijab}
\end{align}\ese%
\bse\label{dynamical-extended-2}\begin{align}
\{Q_{\dot{\alpha}\beta},Q^{\dagger\dot{\rho}\lambda}\} &=
\delta_{\dot{\alpha}}^{\dot{\rho}}\delta_{\beta}^{\lambda} (\H-Z)
+\mu(i\sigma^{ij})_{\dot{\alpha}}^{\dot{\rho}}
\delta_{\beta}^{\lambda}{\bf J}_{ij} +
\mu\delta_{\dot{\alpha}}^{\dot{\rho}}
(i\sigma^{ab})_{\beta}^{\lambda}{\bf J}_{ab} \cr &+
\mu\delta_{\dot{\alpha}}^{\dot{\rho}}(i\sigma^{ab})_{\beta}^{\lambda}
\R_{ab}-\mu(i\sigma^{ij})_{\dot{\alpha}}^{\dot{\rho}}\delta_{\beta}^{\lambda}
\R_{ij}-\mu(i\sigma^{ij})_{\dot{\alpha}}^{\dot{\rho}}
(i\sigma^{ab})_{\beta}^{\lambda}\R_{ijab}^{-+} \\
\{Q_{\dot{\alpha}\beta},Q_{\dot{\rho}\lambda}\} &=-
\mu\delta_{\dot{\alpha}\dot{\rho}}\delta_{\beta\lambda}(C-i\hat{C})
+4\mu(i\sigma^{ij})_{\dot{\alpha}\dot{\rho}}
(i\sigma^{ab})_{\beta\lambda}\M^{-+}_{ijab}
\end{align}\ese%
\bse\label{dynamical-extension-mixing-1}\begin{align}
\{Q_{\dot{\alpha}\beta},Q^{\dagger\rho\dot{\lambda}}\} &=
\mu(\sigma^{i})_{\dot{\alpha}}^{\rho}(\sigma^{a})_{\beta}^{\dot{\lambda}}
(C_{ia}-i\hat{C}_{ia}) \\
\{Q_{\dot\alpha\beta},Q_{\rho\dot\lambda}\} &=
\mu(\sigma^i)_{\dot\alpha\rho}(\sigma^a)_{\beta\dot\lambda}D_{ia}
\end{align}\ese%
\bse\label{dynamical-extension-mixing-2}\begin{align}
\{Q_{\alpha\dot{\beta}},Q^{\dagger\dot{\rho}\lambda}\} &=
\mu(\sigma^{i})_{\alpha}^{\dot{\rho}}(\sigma^{a})_{\dot{\beta}}^{\lambda}
(C_{ia}+i\hat{C}_{ia}) \\
\{Q_{\alpha\dot\beta},Q_{\dot\rho\lambda}\} &=
\mu(\sigma^i)_{\alpha\dot\rho}(\sigma^a)_{\dot\beta\lambda}D_{ia}
\end{align}\ese%
In the above $\R_{ab}=\R_{-10ab}$, $\R_{ij}=\R_{ij56}$ and%
\be \R_{ijab}^{s_{1}s_{2}}\equiv
(\delta_{ik}\delta_{jl}+\frac{s_{1}}{2}\
\epsilon_{ijkl})(\delta_{ac}\delta_{bd} + \frac{s_{2}}{2}\
\epsilon_{abcd})\R_{klcd}\ee%
where $s_1, s_2$ take $\pm$ values. As we see the components of
$\R^{s_1s_2}_{ijab}$ with $s_1 s_2=-1$ appear in the dynamical
superalgebra.

The $\M_{ijab}$ which has appeared in (\ref{dynamical-extended-1}b)
and (\ref{dynamical-extended-2}b) results from the
$\Z_{\hat{\mu}\hat{\nu}\hat{\rho}ABC}^{++}$ extension by setting \be
\M_{ijab}=\Z_{-1ij5ab}^{++}\ . \ee Recalling the six dimensional
self-duality condition \eqref{6d-selfdual}, these are the only
independent components of $\Z^{++}$ with four $so(4)\times so(4)$
indices. In the $so(4)\times so(4)$ notation $\M_{ijab}$ are not
irreducible and one can still reduce $\M_{ijab}$ to self-dual and
anti-self dual parts: \be \M_{ijab}^{s_{1}s_{2}}\equiv
(\delta_{ik}\delta_{jl}+\frac{s_{1}}{2}\
\epsilon_{ijkl})(\delta_{ac}\delta_{bd} + \frac{s_{2}}{2}\
\epsilon_{abcd})\M_{klcd}\ .
\ee%
where the $s_1s_2=-1$ combinations appear in the dynamical part of
the superalgebra.

As we can see  the (\ref{dynamical-extended-1}a,b) and
(\ref{dynamical-extended-2}a,b) are the maximal possible extensions
of the $psu(2|2)$ algebras. One should, however, note that as a
result of the maximal extension the dynamical part of the extended
plane-wave superalgebra is \textit{not} a direct product of the two
extended $psu(2|2)$ factors and they mix through
(\ref{dynamical-extension-mixing-1}),
\eqref{dynamical-extension-mixing-2}.

We would also like to stress that the extensions $C$, $\hat C$, and
$\M_{ijab}$, similarly to their counterparts in the extended
$psu(2|2,4)$ algebra $\Z_{A\hat\mu}$ and  $
\Z_{\hat{\mu}\hat{\nu}\hat{\rho}ABC}^{++}$, are complex valued.

\subsection{Extensions of the kinematical superalgebra}
Having used  the above decomposition and scalings,
we obtain the extended kinematical part of the plane-wave superalgebra%
\bse\label{kinematical-extension-1}\begin{align}
\{q_{\alpha\beta},q^{\dagger\rho\lambda}\} &=
2\delta_{\alpha}^{\rho}\delta_{\beta}^{\lambda}(P^{+}+z) +
(i\sigma^{ij})_{\alpha}^{\rho}\ (i\sigma^{ab})_{\beta}^{\lambda}\
r_{ijab}^{++} \\ \{q_{\alpha\beta},q_{\rho\lambda}\} &=
\frac{-1}{\mu}\delta_{\alpha\rho}\delta_{\beta\lambda}(c-i\hat{c})+
(i\sigma^{ij})_{\alpha\rho}\ (i\sigma^{ab})_{\beta\lambda}\
m_{ijab}^{++} \end{align}\ese%
\bse\label{kinematical-extension-2}\begin{align}
\{q_{\dot{\alpha}\dot{\beta}},q^{\dagger\dot{\rho}\dot{\lambda}}\}
&=
2\delta_{\dot{\alpha}}^{\dot{\rho}}\delta_{\dot{\beta}}^{\dot{\lambda}}
(P^{+}-z)-(i\sigma^{ij})_{\dot{\alpha}}^{\dot{\rho}}\
(i\sigma^{ab})_{\dot{\beta}}^{\dot{\lambda}}\ r_{ijab}^{--}
\\ \{q_{\dot{\alpha}\dot{\beta}},q_{\dot{\rho}\dot{\lambda}}\} &=
\frac{-1}{\mu}\delta_{\dot{\alpha}\dot{\rho}}\delta_{\dot{\beta}
\dot{\lambda}}(c+i\hat{c})+(i\sigma^{ij})_{\dot\alpha\dot\rho}\
(i\sigma^{ab})_{\dot\beta\dot\lambda}\ m_{ijab}^{--}.
\end{align}\ese%
\bse\label{mixed-extension-1}\begin{align}
\{q_{\alpha\beta},q^{\dagger\dot{\rho}\dot{\lambda}}\} &=
\frac{1}{\mu}(\sigma^{i})_{\alpha}^{\dot{\rho}}(\sigma^{a})_{\beta}^{\dot{\lambda}}
(c_{ia}-i\hat{c}_{ia}) \\
\{q_{\alpha\beta},q_{\dot{\rho}\dot{\lambda}}\} &=
\mu(\sigma^i)_{\alpha\dot\rho}(\sigma^a)_{\beta\dot\lambda}d_{ia}
\end{align}\ese%
\bse\label{mixed-extension-2}\begin{align}
\{q_{\dot{\alpha}\dot{\beta}},q^{\dagger\rho\lambda}\} &=
\frac{1}{\mu}(\sigma^{i})_{\dot{\alpha}}^{\rho}(\sigma^{a})_{\dot{\beta}}^{\lambda}
(c_{ia}+i\hat{c}_{ia}) \\
\{q_{\dot{\alpha}\dot{\beta}},q_{\rho\lambda}\} &=
\mu(\sigma^i)_{\dot\alpha\rho}(\sigma^a)_{\dot\beta\lambda}d_{ia}
\end{align}\ese%
In the above $r_{ijab}$ and $m_{ijab}$ are obtained from
$\R^{s_1s_2}_{ijab}$ and $\M^{s_1s_2}_{ijab}$ after the following
rescalings: \be r^{s_1s_2}_{ijab}=\frac{1}{\mu
R^2}\R^{s_1s_2}_{ijab},\ \ m^{s_1s_2}_{ijab}=\frac{1}{\mu
R^2}\M^{s_1s_2}_{ijab},\ \ \ \ s_1s_2=+1 \ee

There are of course non-vanishing kinematical-dynamical
anticommutators which we do not present here and can be worked out
in a similar way.
\section{Extensions from Tiny Graviton Matrix Theory}%
In the previous section through the Penrose limit process we
obtained the plane-wave superalgebra and its most general extension.
In this section we try to realize this superalgebra as the symmetry
of a physical theory. Obviously this physical theory should be
related to string theory (and possibly to its extended objects) on
the plane-wave background. A realization of the unextended
plane-wave supersymmetry algebra
\eqref{qq-anticommutator}-\eqref{QQ-anticommutator} algerba in terms
of the worldsheet coordinates of the strings has been the guiding
principle in obtaining the light-cone plane-wave string field theory
(see section 8 of \cite{plane-wave-review} for a detailed review).
In this formulation, being a perturbative string theory formulation,
however, the extensions which correspond to extended objects (BPS
branes) are absent.

In \cite{TGMT} a non-perturbative formulation of plane-wave string
theory was proposed, according which the DLCQ of type IIB on the
plane-wave background has Matrix quantum mechanics (a $0+1$
dimensional $U(J)$ gauge theory) description, the tiny graviton
matrix theory (TGMT). For convenience and completeness, a very short
introduction to the tiny graviton Matrix theory is given in the
appendix \ref{review}.

In this setup, being a non-perturbative description, one would
expect the extensions to appear naturally. In this section our aim
is to obtain explicit expressions for some of the extensions (brane
charges) in terms of the Matrix degrees of freedom. As a DLCQ
description, we would primarily be interested in the extension to
the dynamical part of the supersymmetry algebra.
The extensions, being correlated with physical observables, should
appear as gauge invariant combination of the $J\times J$ matrices of
the TGMT, and as in the BFSS case \cite{BSS, Taylor}, are
generically in the form of trace of commutators. \footnote{For the
DLCQ of M-theory on the eleven dimensional plane-wave there also
exists a Matrix theory, usually known as BMN or plane-wave matrix
model \cite{BMN}.  The extensions of the eleven dimensional
superalgebra in the context of BMN matrix has been discussed in
\cite{Hyun, Park}.} In the TGMT, however, besides the usual
commutators we also have the option of the four brackets (see the
Appendix \ref{review}). Hence, these extensions, which generically
correspond to brane charges, are vanishing for finite size matrices.
(Recall that in Matrix theory the space integration over
world-volume (of total derivatives) is replaced with the trace over
$U(J)$ indices (of commutators).)

\subsection{Computation of extensions in terms of matrices}

Let us start with the expression for the dynamical supercharges given in \cite{TGMT}:%
\be\label{Q-d-und}
\begin{split}
Q_{\dot{\alpha}\beta} = \sqrt{\frac{R_{-}}{2}}\
\Tr&\bigg[(\Pi^{i}-\frac{i\mu}{R_{-}}X^{i})
(\sigma^{i})_{\dot{\alpha}}^{\rho}\theta_{\rho\beta} +
(\Pi^{a}-\frac{i\mu}{R_{-}}X^{a})(\sigma^{a})_{\beta}^{\dot{\rho}}
\theta_{\dot{\alpha}\dot{\rho}} \cr -&
\frac{i}{3!g_{s}}\big(\epsilon^{ijkl}[X^{i},X^{j},X^{k},\L5]
(\sigma^{l})_{\dot{\alpha}}^{\rho}\theta_{\rho\beta} +
\epsilon^{abcd}[X^{a},X^{b},X^{c},\L5](\sigma^{d})_{\beta}^{\dot{\rho}}
\theta_{\dot{\alpha}\dot{\rho}}\big) \cr +&
\frac{1}{2g_{s}}\big([X^{i},X^{a},X^{b},\L5]
(\sigma^{i})_{\dot{\alpha}}^{\rho}(i\sigma^{ab})_{\beta}^{\gamma}\theta_{\rho\gamma}
+ [X^{a},X^{i},X^{j},\L5](\sigma^{a})_{\beta}^{\dot{\gamma}}
(i\sigma^{ij})_{\dot{\alpha}}^{\dot{\rho}}\theta_{\dot{\rho}\dot{\gamma}}\big)\bigg]
\end{split}
\ee%
\be\label{Q-und-d}
\begin{split} Q_{\alpha\dot\beta} = \sqrt{\frac{R_{-}}{2}}\
\Tr&\bigg[(\Pi^{i}-\frac{i\mu}{R_{-}}X^{i})
(\sigma^{i})_{\alpha}^{\dot\rho}\theta_{\dot\rho\dot\beta} +
(\Pi^{a}-\frac{i\mu}{R_{-}}X^{a})(\sigma^{a})_{\dot\beta}^{\rho}
\theta_{\alpha\rho} \cr -&
\frac{i}{3!g_{s}}\big(\epsilon^{ijkl}[X^{i},X^{j},X^{k},\L5]
(\sigma^{l})_{\alpha}^{\dot\rho}\theta_{\dot\rho\dot\beta} +
\epsilon^{abcd}[X^{a},X^{b},X^{c},\L5](\sigma^{d})_{\dot\beta}^{\rho}
\theta_{\alpha\rho}\big) \cr +&
\frac{1}{2g_{s}}\big([X^{i},X^{a},X^{b},\L5]
(\sigma^{i})_{\alpha}^{\dot\rho}(i\sigma^{ab})_{\dot\beta}^{\dot\gamma}
\theta_{\dot\rho\dot\gamma} +
[X^{a},X^{i},X^{j},\L5](\sigma^{a})_{\dot\beta}^{\gamma}
(i\sigma^{ij})_{\alpha}^{\rho}\theta_{\rho\gamma}\big)\bigg]
\end{split}\ee%
In our conventions the complex  conjugate $^\dagger$ just rises the
lower
indices and vice versa. Note that supercharges are trace of corresponding densities%
\be Q_{\alpha\dot\beta} = \Tr (Q_{\alpha\dot\beta})^{p}_{\
q}=(Q_{\alpha\dot\beta})^{p}_{\ p} \quad,\quad Q_{\dot\alpha\beta} =
\Tr (Q_{\dot\alpha\beta})^{p}_{\ q}=(Q_{\alpha\dot\beta})^{p}_{\ p}
\ee%
where $p,q, r, s=1,2,\cdots , J$ are matrix indices.

In \cite{TGMT}, the above expressions were proposed on the basis
that their anticommutator produced the correct Hamiltonian (among
other operators). Moreover, they have the right behavior under the
symmetries, especially under the $\mathbb{Z}_2$ symmetry which
exchanges $X^i\leftrightarrow X^a$ and the $Q_{\alpha\dot\beta}$ and
$Q_{\dot\alpha\beta}$. To find the explicit form of (some of) the
extensions we perform a careful computation of various
anticommutators of the above dynamical supercharges. (A similar
calculation can be carried out for kinematical and mixed
supercharges.) For the computation we use the following basic
operatorial (to be compared with matrix) commutation relations:
\bea\label{basic-commutators}
 [X^I_{pq}, \Pi^J_{rs}] &=& i\delta^{IJ}\ \delta_{ps}\delta_{qr}
\cr \{(\theta^{\dagger \alpha\beta})_{pq},
(\theta_{\rho\gamma})_{rs}\} &=& \delta^{\alpha}_{\rho}
\delta^{\beta}_{\gamma}\ \delta_{ps}\delta_{qr} \\
\{(\theta^{\dagger \dot\alpha\dot\beta})_{pq},
(\theta_{\dot\rho\dot\gamma})_{rs}\} &=&
\delta^{\dot\alpha}_{\dot\rho} \delta^{\dot\beta}_{\dot\gamma}\
\delta_{ps}\delta_{qr}\nonumber
\eea%
After some straightforward, but lengthy, algebra one can check that
\be\begin{split}
\{Q_{\dot{\alpha}\beta},Q^{\dagger\dot{\rho}\lambda}\} &=
\delta_{\dot{\alpha}}^{\dot{\rho}} \delta_{\beta}^{\lambda}\H +
\mu(i\sigma^{ij})_{\dot{\alpha}}^{\dot{\rho}}
\delta_{\beta}^{\lambda}{\bf J}_{ij} +
\mu\delta_{\dot{\alpha}}^{\dot{\rho}}
(i\sigma^{ab})_{\beta}^{\lambda}{\bf J}_{ab} -
\mu(i\sigma^{ij})_{\dot{\alpha}}^{\dot{\rho}}(i\sigma^{ab})_{\beta}^{\lambda}
\R_{ijab}\end{split}\ee%
and similarly for
$\{Q_{{\alpha}\dot\beta},Q^{\dagger{\rho}\dot\lambda}\}$, where
$\H$, ${\bf J}_{ij}$ and ${\bf J}_{ab}$ are given in
\eqref{Hamiltonian}, \eqref{Jij} and \eqref{Jab} of the Appendix
\ref{review}. {}From the above form of the supercharges it is
readily seen that \be\label{Rijab-Matrix} \R_{ijab} =
\frac{1}{g_s}\Tr\big([X^{i},X^{j},X^{a},X^{b}]\L5\big)
\ee%

Next, one may show that \be\label{QQ-Matrix}
\begin{split}
\{Q_{\dot{\alpha}\beta},Q_{\dot{\rho}\lambda}\} = 0\ & ,\ \ \ \ \ \
\{Q_{{\alpha}\dot\beta},Q_{{\rho}\dot\lambda}\} = 0\ ,\cr
\{Q_{\dot{\alpha}\beta} & ,Q_{{\rho}\dot \lambda}\} = 0\ .
\end{split}\ee%
from which we learn that in the TGMT defined via the Hamiltonian
\eqref{Hamiltonian} \be\label{C-hatC-M} C=\hat C=0\ ,\ \ \ \
D_{ia}=0\ ,\ \ \ \ \ \M_{ijab}=0\ . \ee And finally one can, in a
similar way show that \be\begin{split}
\{Q_{\dot{\alpha}\beta},Q^{\dagger\rho\dot{\lambda}}\} &\equiv \mu
(\sigma^{i})_{\dot{\alpha}}^{\rho}(\sigma^{a})_{\beta}^{\dot{\lambda}}
(C_{ia}-i\hat{C}_{ia}) \cr \end{split}\ee%
{}From the above and recalling that $\hat C_{ia}$ and  $C_{ia}$ are
both hermitian (note \eqref{QQdagger-extended} and
(\ref{Penrose-on-extensions}a,b)) one can read the explicit form of
$C_{ia}$ and  $\hat C_{ia}$%
\be\begin{split} C^{ia} = \frac{R_-}{\mu}\Tr\bigg[&P^iP^a
-\left(\frac{1}{2g_s}\right)^2
\epsilon^{abcd}\epsilon^{ijkl}[X^{j},X^{b},X^{c},{\cal L}_5]
[X^{d},X^{k},X^{l},{\cal L}_5]\cr & +
\Big(\frac{\mu}{R_{-}}X^{i}+\frac{1}{3!g_s}\epsilon^{ijkl}[X^{j},X^{k},X^{l},{\cal
L}_5]\Big)\Big(\frac{\mu}{R_{-}}X^{a}+\frac{1}{3!g_s}\epsilon^{abcd}[X^{b},X^{c},X^{d},{\cal
L}_5]\Big)\bigg]\end{split}\ee%
\be \hat C^{ia} =
\frac{R_-}{\mu}\frac{1}{2g_s}\Tr\Big(\epsilon^{ijkl}P^j[X^a,X^k,X^l,\L5]
+ \epsilon^{abcd}P^c[X^i,X^c,X^d,\L5]\Big) \ee%

We would like to stress that in the above computations we have
assumed the finite $J$ condition and have set $\Tr([A,B])$ and
$\Tr([A,B,C,D])$ equal to zero. Moreover, in the above computations
we have explicitly used the Gauss law constraint \eqref{Gauss-law}.

\subsection{Physical interpretation of the extensions}

Having computed the extensions using the expression of the
supercharges \eqref{Q-und-d} and \eqref{Q-d-und}, here we discuss
some of their physical aspects:

$\bullet$ As we can explicitly see from the above computations some
of the extensions are found to be zero (for finite size matrices).
Noting the equations \eqref{Penrose-on-extensions} it is seen that
all the non-vanishing extensions are coming from the
$\R_{AB\hat\mu\hat\nu}$ extension of the original extended
$psu(2,2|4)$ algebra ({\it cf.} \eqref{QQdagger-extended}). As we
discussed in section 2 this extension corresponds to three brane
charges. The extensions $\chi,\ \Z_{A\hat\mu},\
\Z_{ABC\hat\mu\hat\nu\hat\rho}^{++}$ are all vanishing, indicating
that in the current form of the TGMT  -1, 1, and 5 branes are not
present.

The above was of course expected recalling that the TGMT was
obtained by the quantization (discretization) of a three brane in
the plane-wave background and that in the corresponding Born-Infeld
action only the contribution of the RR four-form to the Wess-Zumino
terms were included \cite{TGMT}.

$\bullet$ The extensions account for the three brane RR {\it dipole}
moments, as well as charges.

In the usual superPoincare' algebras the $p$-form (central)
extensions are identified with the RR charge of {\it flat}
D$p$-branes. One of the consequences of this identification is that
flat branes in a flat  Minkowski background are 1/2 BPS objects. In
a non-flat background, such as $AdS_5\times S^5$ or the ten
dimensional plane-wave, the half BPS objects are not flat branes.
These are spherical three branes, the giant gravitons \cite{Giants}.
The three brane giant gravitons, being spherical, do not carry a net
RR charge of the (selfdual) RR four-form. They, however, carry
dipole moment of the four-form. For the spherical branes one can
compute the dipole moment corresponding to the brane. This dipole
moment is then naturally a five-form. Noting that each small element
on the spherical brane locally behaves as a three brane which
carries a unit of the corresponding RR charge density, one concludes
that the RR dipole moment should be proportional to the volume form
of the embedding space. To make the above more quantitative, recall
that in a three brane action the coupling of a three brane to
external RR four form is of the form
\[
S_{C_4}=\int d\tau d^3\sigma \epsilon_{rsp}
C_{\mu\nu\rho\alpha}\partial_{\tau}X^\mu
\partial_{\sigma_r}X^\nu \partial_{\sigma_p}X^\rho \partial_{\sigma_s}X^\alpha\ ,
\]
where the Greek indices run over $0,\cdots, 9$, $X^\mu$ are the
embedding coordinates of the brane and $r,p,s$ indices run over
$1,2,3$. If we fix the light-cone gauge, assuming that our $C_4$
field  has a constant field strength $F^{(5)}$, we arrive at
\cite{TGMT, hedgehog} \be\label{dipole}
\begin{split}
S_{C_4}&=\int d\tau d^3\sigma \epsilon_{rsp}\  F^{(5)}_{+IJKL}\
X^I\partial_{\sigma_r}X^J \partial_{\sigma_p}X^K
\partial_{\sigma_s}X^L\cr &= \int d\tau d^3\sigma\ F^{(5)}_{+IJKL}\
X^I \{X^J,X^K, X^L\}
\end{split}
\ee where $I,J,K,L=1,2,\cdots 8$ are the transverse light-cone
coordinates and the $\{\cdot, \cdot, \cdot\}$ is the Nambu
3-bracket. From the above it is readily seen that the RR dipole
moment fiveform of the brane is
\[
d^{(5)}_{+IJKL}=\int d^3\sigma\ X^I \{X^J,X^K, X^L\}
\]
Upon the quantization prescription discussed in \cite{TGMT} the
fiveform dipole moment in the TGMT takes the form
\be\label{dipole-TGMT} d^{(5)}_{IJKL}\mid_{_{TGMT}}= \Tr
(X^I[X^J,X^K,X^L,\L5])= \Tr ([X^I,X^J,X^K,X^L]\L5) \ee We should
emphasize that the above dipole moment expression should only be
used for the transverse branes, those which contain the (light-cone)
time direction $X^+$, but not the $X^-$.

The above arguments is not limited to three branes and can be
repeated for any kind of D-brane alike. For the {\it even}
dimensional branes this has been previously discussed in the
literature and known as the Myers dielectric effect \cite{Myers}.

Before moving to some specific examples we would like to briefly
discuss the matrix $\L5$ which has appeared in the dipole moment
expression. As noted in \cite{TGMT} in order to pass from the
classical Nambu 3-brackets (and in general Nambu odd-brackets) to
their quantum version we need to introduce an appropriate operator
(or matrix). For our purposes this \textit{fixed} matrix was called
$\L5$. In \cite{half-bps} a more precise definition of the $\L5$ was
given and its physical meaning  was uncovered: $\L5$ is the
reminiscent of the 11$^{th}$ circle. From the charge analysis
appeared above and as can be seen from the second equality in
\eqref{dipole-TGMT}, $\L5$ is necessary to obtain a non-zero
\textit{dipole} moment. In other words, as discussed in
\cite{half-bps},  by definition $\L5$ is a traceless $J\times J$
matrix which squares to one. Hence, in the   diagonal basis it is
has $J/2$ plus one and $J/2$ minus one eigenvalues and intuitively
one may think of the positive (negative) eigenvalues corresponding
to upper (lower) half of the three sphere. These two semi-spheres
have equal but opposite RR charges while their contribution to the
RR dipole moment is summed up. This fact is exactly reflected in the
form of $\L5$.

Depending on $X^I$ to take value $X^i$ or $X^a$ the above dipole
moment can be decomposed into various irreducible representations of
$so(4)\times so(4)$. If all the $X$'s appearing in $d^{(5)}$ are of
the form of $X^i$ or $X^a$, then the dipole moment $d^{(5)}$ is
singlet of both of the $so(4)$'s (note that $\epsilon_{ijkl}$ or
$\epsilon_{abcd}$ are singlets of both $so(4)$'s). In this case, for
the 1/2 BPS spherical branes discussed in detail in \cite{half-bps},
one can explicitly compute the dipole moment. For the 1/2 BPS
spherical solutions
\[
[X^i,X^j,X^k,\L5]=-\frac{\mu g_s}{R_-} \epsilon_{ijkl} X^l,
\]
and hence \be\label{d5-1/2BPS}
\begin{split}
d^{(5)}_{ijkl}\mid_{\;_{1/2\ BPS}}& =\epsilon_{ijkl}\ \frac{\mu
g_s}{R_-} \Tr(X^mX_m)\cr & =\epsilon_{ijkl}\ \left(\frac{\mu
g_s}{R_-}\right)^2 \sum_{i=1}^k J_i^2\ .
\end{split}
\ee In the above we have considered a generic concentric
configuration of $k$ giant gravitons of radius $J_i$, where $\sum
J_i=J$ \cite{half-bps}. For a single giant solution, the $X=J$
vacuum, where $k=1$
\[
d^{(5)}_{_{X=J}}=\epsilon_{ijkl} \ \left(\frac{\mu J}{R_-}
g_s\right)^2 =\epsilon_{ijkl} R_{giant}^4\ ,
\]
where $R_{giant}^2=\mu p^+ g_s$ with $p^+=J/R_-$, is the radius of
the giant graviton in the string units. This value of $d^{(5)}$ is
the maximum value \eqref{d5-1/2BPS} can take. For the $X=0$ vacuum,
where $k=J$ or $J_i=1$, we have the minimum possible dipole moment
which is equal to
\[
d^{(5)}_{_{X=0}}=\epsilon_{ijkl} \ \left(\frac{\mu
g_s}{R_-}\right)^2 J =\epsilon_{ijkl}\ l_{tiny}^4 J\ .
\]
where $l_{tiny}$ is the size of the tiny three brane gravitons
\cite{TGMT, half-bps}. The above is physically expected noting that
it is the dipole moment of a single brane times their number $J$.
The result that the dipole moments of branes are additive is generic
and can be seen from \eqref{d5-1/2BPS}, recalling that the size of a
giant which carries $J_i$ units of the light-cone momentum is
$R^2_i=\mu J_i g_s/R_-$.

It is notable that the dipole moment energy for these configurations
is canceled out with the spherical brane tension \cite{half-bps}
such that these spherical solutions are all eigenstates of  the
light-cone Hamiltonian with eigenvalue zero.

Among the components of $d^{(5)}_{IJKL}$ which mix $X^i$ and $X^a$
directions only those with two $X^i$ and two $X^a$ appear as the
extensions of the superalgebra in the form of $\R_{ijab}$. These are
the RR dipole moment of the (topologically spherical) three branes
embedded in both $X^i$ and $X^a$ direction. A detailed analysis of
these states will be given in \cite{bpscfg}.

Finally we have the $C_{ia}$ and $\hat C_{ia}$ extensions which
correspond to \textit{longitudinal} flat three branes containing
$X^i$ and $X^a$ as well as $X^+$ and $X^-$ directions. This can be
seen tracing back the origin of the $C_{ia}$ extension to before the
Penrose limit and that $C_{ia}$ is related to $\R_{0i5a}$ component
of the $\R$ extension. (Note that $x_{-1},\ x_0$ and $x_5,\ x_6$
combine to give  $x^+,\ x^-$. The two other directions are basically
transverse to the $AdS_5\times S^5$ and do not appear at all.)

\section{Discussion and Outlook}%

In this work we have considered the most general extension of the
ten dimensional plane-wave superalgebra. To obtain that we started
with the corresponding superalgebra, the maximally extended
$psu(2,2|4)$ algebra, and took the Penrose limit over that. Under
this procedure the algebra naturally decomposes into the kinematical
and dynamical parts. The dynamical part is of the form of the
maximally extended $ps(2|2)\times psu(2|2)\times u(1)_{\bf H}$, as
one would have expected if we started directly from the plane-wave
algebra and studied its maximal extension. In other words, extending
the algebra commutes with process of taking the Penrose limit. In
this way, however, we have the virtue of a direct relation of the
extension of the plane-wave to branes and their charges and whether
these branes are longitudinal or transverse. We have shown that only
the extensions corresponding to three branes appear in the dynamical
part of the superalgebra appear in the present form of the TGMT;
these three branes can however be transverse or longitudinal.

Given the form structure of the extensions and their $so(4)\times
so(4)$ representation it is straightforward to see that the BPS
branes allowed through our extended plane-wave superalgebra are in
one-to-one correspondence with the BPS branes expected from string
theory discussion of \cite{planewave-brane}.

We gave an explicit realization of the supercharges, and in
particular the dynamical ones, in the context of the Tiny Graviton
Matrix Theory (TGMT). We discussed that the TGMT on the pure
plane-wave background naturally admits and contains the central
extension corresponding to spherical transverse three branes or flat
longitudinal branes. An interesting outcome of our Matrix theory
analysis is that the extensions can be related to the RR
\textit{dipole} moment of the branes as well as the usually
discussed charges. In fact if the BPS brane configuration does not
carry a net RR charge, as is the case with the giant gravitons in
the $AdS$ or plane-wave background, then the lowest moment, which in
this case is the dipole moment, can appear as the extension in the
algebra. We made this fact  manifest  in the TGMT. The fact that the
dipole moments can also appear in the superalgebras as the
extensions can have a profound effect in counting of the microstates
of BPS blackholes in non-flat background. In particular it implies
that there could possibly be some modifications to the no hair
theorem, especially in the context of higher dimensional blackholes
and blackholes in non-flat backgrounds. The possibility of
appearance of the dipole moments in the blackhole thermodynamics has
been recently discussed \cite{dipole-counting}. In light of the
above discussions one may then try to account for the dipole moment
effects in the blackhole thermodynamics through the superalgebra and
its representations and classification of the dipole-charged BPS
objects.

One of the interesting points which appears in studying the
extensions of superalgebras are the concept of ``fermionic brane
charges'' which was first noted in \cite{M-algebra} and further
discussed in \cite{Peeters}. It would be nice to study these
fermionic brane charges further in the context of the TGMT or the
BMN Matrix model \cite{BMN}. These fermionic charges should show up
when we compute commutator of the extensions with the supercharges.
In the TGMT where  we have given explicit form of both of these in
terms of $J\times J$ matrices it is then straightforward to carry
out such computations.

Having obtained the explicit form of the extended superalgebra in
terms of the TGMT one can now extend the work of \cite{half-bps} to
less supersymmetric configurations  \cite{bpscfg}. With the
formulation at hand and with finite size matrices one can only
analyze compact three brane configurations. For the infinite extent
branes we should take the infinite $J$ limit. In order to include
other branes, e.g. D-strings or D5/NS5-branes we need to extend the
TGMT by adding the appropriate terms which account for these
objects. This is very similar to what is done for the BFSS matrix
model \cite{R-T}.

In \cite{half-bps}, based on the results in the 1/2 BPS sector, we
proposed a triality between the TGMT, type IIB string theory and the
$\N=4$ SYM gauge theory. The latter two are of course related via
the usual AdS/CFT. In order to provide further support for the
proposal we need to find the explicit representation and
manifestation of the fully extended $psu(2,2|4)$ superalgebra in
terms of the (gauge invariant) operators of the $\N=4$ four
dimensional  gauge theory. Moreover, from the TGMT side we need to
have a classification of the less BPS configurations of the TGMT to
match it against similar configurations in the gravity or dual gauge
theory side.

\appendix
\section{Fermion Notations}\label{Notations}%
The supersymmetry algebra of $AdS_5\times S^5$, in the context of
Kac-Nahm classification of Lie superalgebras of classical type, is
$PSU(2,2|4)$, which is also superalgebra of $D=4, \N=4$
superconformal field theory. This superalgebra may be represented
using 4,10 or 10+2 dimensional notations according to taste. We
shall present this superalgebra in the $so(4,2)\oplus so(6)$ and
coset $so(4,1)\oplus so(5)$ basis.
\subsection{Twelve dimensional fermions in $SO(4,2)\times SO(6)$ notations}
In order to make the isometry of $AdS_5\times S^5$ manifest, it is
convenient to employ (10+2)-dimensional notation. 12-dimensional
gamma
matrices $\G^{\hat m}$ satisfy%
\be \{\G^{\hat m},\G^{\hat n}\} = 2g^{\hat m\hat n}
\ee%
with $g^{\hat m\hat n} = diag(--++++++++++)$ and $\hat m =
-1,0,1,\dots,10$. They can be represented in terms of six
dimensional gamma matrices $\Gamma^{\hat\mu}$ and $\Gamma^{A}$ with
appropriate signature corresponding to $spin(4,2)$ and $spin(6)$
spin groups, as%
\be \G^{\hat\mu} = \Gamma^{\hat\mu}\otimes\Gamma^7,\quad \G^{A} =
{\bf 1_8}\otimes\Gamma^{A} \ee
where six dimensional chirality matrix is%
\be \Gamma^7 = i\Gamma^{-1}\dots\Gamma^4 = i\Gamma^5\dots\Gamma^{10}
= \left(\begin{array}{cc} {\bf 1_4} & 0 \\ 0 & {\bf -1_4}
\end{array}\right)
\ee%
with the following condition for six dimensional $8\by 8$ gamma
matrices%
\be \{\Gamma^{\hat\mu},\Gamma^{\hat\nu}\} =
2\eta^{\hat\mu\hat\nu},\quad
\{\Gamma^{A},\Gamma^{B}\} = 2\delta^{AB} \ee%
For the sake of space we do everything here for $SO(6)$, similar
things would happen for $SO(4,2)$ just substitute 5 with -1 and 6
with 0, and the corresponding metric signature.

In six dimensions we are dealing with 8 component Dirac fermions,
$\theta_\mu$. These Dirac spinors can be decomposed into two Weyl
spinors $\theta_I$ and $\theta_{\dot I}$ where $I,\dot I = 1,2,3,4$
are fundamental, antifundamental indices. $\Gamma$ matrices can be
decomposed into $\Gamma^\pm$ and $\Gamma^i$, and with a convenient
choice of basis they can be written as%
\be \Gamma^+ = i\left(\begin{array}{cc}  0 & \sqrt 2 \\  0 & 0 \\
\end{array} \right),\ \Gamma^- = i\left(\begin{array}{cc}  0 & 0 \\
\sqrt 2 & 0 \\ \end{array} \right),\ \Gamma^i =
\left(\begin{array}{cc} \gamma^i & 0 \\ 0 & -\gamma^i \\ \end{array}
\right),\ \Gamma^7 = \left(\begin{array}{cc}  \gamma^5 & 0 \\ 0 & -\gamma^5 \\
\end{array} \right) \ee%

Now $\gamma^i$'s are $4\by 4$ matrices satisfying
$\{\gamma^i,\gamma^j\} = 2\delta^{ij}$ and act on the Weyl spinors.
Now $SO(6)$ Weyl spinors can be seen as Dirac spinors of $SO(4)$ and
in turn can be decomposed into two, 2 component Weyl spinors%
\be \theta_I \rightarrow (\theta_\alpha,\ \theta_{\dot\alpha}) \ee%
where $\alpha,\dot\alpha = 1,2$. The $\gamma_{4\by 4}$ matrices can
also be reduced to $2\by 2$ representations,
$\sigma^i_{\alpha\dot\alpha}$ and $\bar\sigma^i_{\dot\alpha\alpha}$
in a convenient way, to act on Weyl spinors%
\be (\gamma^i)_{I\dot J} = \left(\begin{array}{cc}
0 & (\sigma^i)_{\alpha\dot\beta} \\ (\bar\sigma^i)_{\dot\alpha\beta} & 0 \\
\end{array}\right) \ee%

The $SO(4,2)\times SO(6)$ fermions carry spinorial indices of both
of the groups. Because of 10-dim chirality and the fact that we
choose positive sign for the self-dual five-form flux, out of four
different possibilities, for the $AdS_5\times S^5$ we only need
fermions with the same $SO(4,2)$ and $S(6)$ chirality (see Appendix
B of \cite{plane-wave-review}), \ie $\theta_{I\hat J} \equiv
\theta_I \otimes \theta'_{\hat J}$, and using the above
decompositions it decomposed as $\theta_{I\hat J} \rightarrow
(\theta_{\alpha\beta}, \theta_{\alpha\dot\beta},
\theta_{\dot\alpha\beta}, \theta_{\dot\alpha\dot\beta})$.

\subsection{Ten dimensional fermions in $SO(4)\times SO(4)$ notation}\label{A2}
In this part we present the superalgebra in coset $SO(9,1)\sim
SO(4,1)\oplus SO(5)$ basis. Then we go to onshell $SO(8)$
representations, and finally we give $SO(4)\times SO(4)$
representations, appropriate to manifest isometries of plane-wave.

10-dimensional $32\by 32$ gamma matrices $\Gamma^{\hat a}$ of
$spin(9,1)$ which respect Clifford algebra%
\be \{\Gamma^{\hat a},\Gamma^{\hat b}\} = 2\eta^{\hat a\hat b} \ee%
can be decompose in terms of $4\by 4$ gamma matrices of $spin(4,1)$
and $spin(5)$ as%
\be \Gamma^a = \gamma^a\otimes {\bf 1}\otimes \sigma_1,\quad
\Gamma^{a'} = {\bf 1}\otimes\gamma^{a'}\otimes \sigma_2 \ee%
Where each of them also satisfy%
\be \{\gamma^a,\gamma^b\} = \eta^{ab},\quad
\{\gamma^{a'},\gamma^{b'}\} = \delta^{a'b'} \ee%
In this convention, $\hat a=0,1,\dots,9$ and $a=0,1,\dots,4$ and
$a'=5,6,\dots,9$. It may be interesting to note the 32 component
10-dimensional positive chirality spinor and
negative chirality supercharges are decomposed as%
\be \theta^{\hat\alpha} =
\theta^\alpha\otimes\theta^{\alpha'}\otimes\l(^1_0\r),\quad
Q^{\hat\alpha} = Q^\alpha\otimes Q^{\alpha'}\otimes\l(^{\ 0}_{-1}\r)
\ee%

A convenient choice of basis for $32\by 32$ Dirac matric would be%
\be \Gamma^+ = i\left(\begin{array}{cc}  0 & \sqrt 2 \\  0 & 0 \\
\end{array} \right),\ \Gamma^- = i\left(\begin{array}{cc}  0 & 0 \\
\sqrt 2 & 0 \\ \end{array} \right),\ \Gamma^{\hat i} =
\left(\begin{array}{cc}
\gamma^{\hat i} & 0 \\ 0 & -\gamma^{\hat i} \\
\end{array} \right),\ \Gamma^{11} = \left(\begin{array}{cc}  \gamma^9 & 0 \\ 0 & -\gamma^9 \\
\end{array} \right) \ee%
where $\gamma^{\hat i}$ satisfy $\{\gamma^{\hat i},\gamma^{\hat j}\}
= 2\delta^{\hat i\hat j}$ where $\delta$ is the metric on the
transverse space with $\hat i=1,2,\dots,8$. Dirac spinor in 10
dimensions, $\theta_{\hat\alpha}$, has 32 complex components. First
we impose Majorana reality condition $\theta = \theta^\dagger$, so
we remain with 32 real components. Then we demand Majorana spinors
satisfy on-shell condition%
\be \Gamma^\pm\theta^\pm_{\hat\alpha} = 0 \ee%
with the above decomposition for gamma matrices it can be easily
seen that%
\be \theta^+_{\hat\alpha} = \l(\begin{array}{c} \theta^+_{16} \\ 0
\end{array} \r),\qquad \theta^-_{\hat\alpha} = \l(\begin{array}{c} 0 \\ \theta^-_{16}
\end{array} \r) \ee%
$\theta^\pm_{16}$ can be thought of as $SO(8)$ Majorana fermion, and
$\gamma^{\hat i}$ as $16\by 16$ real matrices. Furthermore we are
dealing with type IIB string theory, so we demand both fermions have
the same ten dimensional chirality, which is related
to eight dimensional chirality as indicated above%
\be \gamma^9 \theta^\pm_{16} = \pm\theta^\pm_{16} \ee%
with 8-dimensional chirality matrix%
\be \gamma^9 = \left(\begin{array}{cc} {\bf 1_8} & 0 \\ 0 & {\bf -1_8} \\
\end{array}\right) \ee%
the solution is%
\be \theta^+_{16} = \left(\begin{array}{c} \theta^+_8 \\ 0 \\
\end{array}\right),\qquad \theta^-_{16} = \left(\begin{array}{c} 0 \\ \theta^-_8
\end{array}\right) \ee%
which are $SO(8)$ Majorana-Weyl spinors, denoted by ${\bf 8_s}$ and
${\bf 8_c}$. Furthermore $16\by 16$ dimensional gamma matrices
can be reduced to $8\by 8$ representations,%
\be (\gamma^{\hat i})_{16} = \left(\begin{array}{cc}
0 & (\gamma^{\hat i})_8 \\ (\bar\gamma^{\hat i})_8 & 0 \\
\end{array}\right) \ee%

We now transfer to $SO(4)\by SO(4)$ representations. To relate
$SO(4)\by SO(4)$ fermions to those of $SO(8)$, we note that ${\bf
8_s}\ ({\bf 8_c})$ has positive (negative) chirality. $SO(8)$
chirality matrix can be written as%
\be \gamma^9 = \gamma^5\times\gamma^{5'}\ee%
in terms of $SO(4)$ chirality matrices%
\be \gamma^5 = \gamma^{1234},\qquad \gamma^{5'} = \gamma^{5678}
\ee%
As mentioned above, we have two spinors of the same chirality,
either two ${\bf 8_s}$'s or two of ${\bf 8_c}$'s. Let's choose them
to be ${\bf 8_s}$, and out of them we can write a complex ${\bf
8_s}$. It can easily be seen that with the above decomposition for
chirality matrix, for ${\bf 8_s}$ the two $SO(4)$ should have the
same chirality while for ${\bf 8_c}$ they should have the opposite
chirality. Note also that an $SO(4)$ spinor can be decomposed into
two Weyl spinors. Explicitly%
\bea {\bf 8_s} &\rightarrow& \theta_{\alpha\beta'};\
\theta_{\dot\alpha\dot\beta'} \\ {\bf 8_c} &\rightarrow&
\theta_{\alpha\dot\beta'};\ \theta_{\dot\alpha\beta'} \eea%
$8\by 8$ gamma matrices also reduce to%
\be \gamma^i_{a\dot a} = \left(\begin{array}{cc} 0 &
(\sigma^i)_{\alpha\dot\beta}
\delta_{\alpha'}^{\beta'} \\
(\sigma^i)^{\dot\alpha\beta}\delta_{\dot\alpha'}^{\dot\beta'} & 0
\end{array}\right),\quad \gamma^i_{\dot aa} = \left(\begin{array}{cc}
0 & (\sigma^i)_{\alpha\dot\beta}\delta_{\dot\alpha'}^{\dot\beta'} \\
(\sigma^i)^{\dot\alpha\beta}\delta_{\alpha'}^{\beta'} & 0
\end{array}\right)\ee%
\be \gamma^{i'}_{a\dot a} = \left(\begin{array}{cc}
-\delta_\alpha^\beta(\sigma^{i'})_{\alpha'\dot\beta'} & 0 \\
0 & \delta_{\dot\alpha}^{\dot\beta}(\sigma^{i'})^{\dot\alpha'\beta'}
\end{array}\right),\quad \gamma^{i'}_{\dot aa} = \left(\begin{array}{cc}
-\delta_\alpha^\beta(\sigma^i)_{\dot\alpha'\beta'} & 0 \\
0 &
\delta_{\dot\alpha}^{\dot\beta}(\sigma^{i'})^{\dot\alpha'\beta'}\end{array}\right)\ee%
with%
\be (\sigma^i)_{\alpha\dot\alpha} = ({\bf 1},
i\vec\sigma)_{\alpha\dot\alpha},\quad (\sigma^i)_{\dot\alpha\alpha}
= ({\bf 1},-i\vec\sigma)_{\dot\alpha\alpha}
\ee%
For simplicity in the notation one may also drop the primes on the
second $so(4)$ indices, whereby we arrive at the fermionic notations
employed in the main text of this paper.
\section{Tiny Graviton Matrix Theory, A Short
Review}\label{review}%
In this appendix we briefly review the basics of the tiny graviton
matrix theory, TGMT. It is essentially a very short summary of
\cite{TGMT}. The tiny graviton matrix theory proposal is  that the
DLCQ of strings on the $AdS_5\times S^5$ or on the 10 dimensional
plane-wave background in the sector with $J$ units of light-cone
momentum is described by the theory or dynamics of $J$ ``tiny''
(three-brane) gravitons. The action for $J$ tiny gravitons is
obtained as a regularized (quantized) version of D3-brane light-cone
Hamiltonian, as has been carried out in \cite{TGMT}. In other words,
DLCQ of type IIB strings on the plane-wave background is nothing but
a quantized 3-brane theory. Then the statement of the conjecture is:
\begin{quote}
{\it The theory of $J$ tiny three-brane gravitons, which is a $U(J)$
supersymmetric quantum mechanics with the \super\ symmetry, is the
Matrix theory describing the DLCQ of strings on the plane-waves or
on the $AdS_5\times S^5$ in the sector with light-cone momentum
$p^+=J/R_-$, $R_-$ being the light-like compactification radius.}
\end{quote}
Dynamics of the theory is governed by the following Hamiltonian%
\be\label{Hamiltonian}
\begin{split}
{\bf H}= R_-\ \Tr&\bl[ \frac{1}{2}(P_i^2+P_a^2) +
\frac{1}{2}\left(\frac{\mu}{R_-}\right)^2(X_i^2+X_a^2) \cr &+
\frac{1}{2\cdot 3!g_s^2} \left([ X^i , X^j , X^k, {\cal L}_5][ X^i ,
X^j , X^k, {\cal L}_5] + [ X^a , X^b , X^c, {\cal L}_5][ X^a , X^b ,
X^c, {\cal L}_5]\right) \cr &+ \frac{1}{2\cdot 2g_s^2} \left([ X^i ,
X^j , X^a, {\cal L}_5][ X^i , X^j , X^a, {\cal L}_5] + [ X^a , X^b ,
X^i, {\cal L}_5][ X^a , X^b , X^i, {\cal L}_5]\right) \cr &
-\frac{\mu}{3!R_- g_s}\left( \epsilon^{i j k l} X^i [X^j, X^k, X^l,
{\cal L}_5]+ \epsilon^{a b c d} X^a [ X^b, X^c, X^d , {\cal L}_5]
\right)\cr &+\left(\frac{\mu}{R_-}\right) \left(\theta^\dagger
{}^{\alpha \beta} \theta_{\alpha \beta}- \theta_{\dot\alpha
\dot\beta}\theta^\dagger {}^{\dot\alpha \dot\beta}\right)\cr
&+\frac{1}{2g_s}\left( \theta^\dagger {}^{\alpha \beta}
(\sigma^{ij})_\alpha^{\:  \: \delta} [ X^i, X^j, \theta_{\delta
\beta}, {\cal L}_5] + \theta^\dagger {}^{\alpha \beta}
(\sigma^{ab})_\alpha^{ \: \: \delta} \: [ X^a, X^b, \theta_{\delta
\beta}, {\cal L}_5]\right) \cr &-\frac{1}{2g_s}
\left(\theta_{\dot\delta \dot\beta} (\sigma^{ij})_{\dot\alpha}^{ \:
\: \dot\delta} \: [ X^i, X^j, \theta^\dagger {}^{\dot\alpha
\dot\beta}, {\cal L}_5]+ \theta_{\dot\delta \dot\beta}
(\sigma^{ab})_{\dot\alpha}^{\: \: \dot\delta} \: [ X^a, X^b,
\theta^\dagger {}^{\dot\alpha \dot\beta}, {\cal L}_5]\right)\br]\,
\end{split}
\ee%
The $\L5$ is a hermitian $J\times J$ defined by \cite{half-bps}
\be\label{L5} \L5^2={\bf 1}\ ,\ \ \ \  \Tr\L5=0\ . \ee

The  $U(J)$ gauge symmetry of the above Hamiltonian is in fact a
discretized (quantized) form of the spatial diffeomorphisms of the
3-brane. As is evident from the above construction we expect in
$J\to\infty$ limit to recover the diffeomorphisms. One should note
that under the $U(J)$ transformations $\L5$ is also transformaing in
the adjoint representation.

The Hamiltonian can also be obtained from a $0+1$ dimensional $U(J)$
gauge theory Lagrangian, in the temporal gauge. Explicitly, the only
component of the gauge field, $\A0$, has been set to zero. To ensure
the $\A0=0$ gauge condition, all of our physical states must satisfy
the Gauss law constraint arising from equations of motion of $\A0$.
These constraints, which consist of $J^2-1$ independent conditions
are:%
\be\label{Gauss-law}
i[X^i,P^i]+i[X^a,P^a]+\{\theta^{\dagger\alpha\beta},\theta_{\alpha\beta}\}
+\{\theta^{\dagger\dot\alpha\dot\beta},\theta_{\dot\alpha\dot\beta}\}
= 0 \ee%
where $P^I = D_0X^I = \partial_0 X^I + i[\A0,X^I]$ and  all the
fields $X, P, \theta, \A0$ and $\L5$ are $J\by J$ matrices.

The Hamiltonian is proposed to describe type IIB string theory on
the plane-wave with compact $X^-$ direction. The ``string theory
limit'' is then a limit where we decompactify $R_-$, keeping $p^+$
fixed, {\it i.e.}%
\be\label{string-theory-limit} J, R_- \to \infty,
\qquad \mu,\ p^+=J/R_-, g_s\ \ {\rm fixed}\ . \ee%
In fact one can show that in the above string theory limit one can
re-scale $X$'s such that $\mu, p^+$ only appear in the combination
$\mu p^+$. Therefore the only parameters of the continuum theory are
$\mu p^+$ and $g_s$.

The plane-wave is a maximally supersymmetric one, {\it i.e.} it has
32 fermionic isometries which can be arranged into two sets of 16,
the kinematical supercharges, $q$'s, and the dynamical supercharges,
$Q$'s. The former are those which anticommute to light-cone momentum
$P^+$ and the latter anticommute to the light-cone Hamiltonian ${\bf
H}$. Here we show the dynamical part of superalgebra, which can be
identified with \super :
\bea [P^{+}, q_{\alpha\beta}]=0 \quad&,&\quad [P^{+},
q_{\dot\alpha\dot\beta}]=0\ ,\cr [{\bf H}, q_{\alpha\beta}]=-i\mu
q_{\alpha\beta} \quad&,&\quad [{\bf H},
q_{\dot\alpha\dot\beta}]=i\mu q_{\dot\alpha\dot\beta}\ . \eea \bea
[P^{+}, Q_{\alpha\dot\beta}]=0 \quad&,&\quad [P^{+},
Q_{\dot\alpha\beta}]=0 \cr [{\bf H}, Q_{\alpha\dot\beta}]=0
\quad&,&\quad [{\bf H},Q_{\dot\alpha\beta}]=0
\eea%
\be\label{qq} \{q_{\alpha
\beta},q^{\dagger\rho\lambda}\}=2P^+\delta_{\alpha}^{\ \rho}
\delta_{\beta}^{\ \lambda}\  , \ \ \ \{q_{\alpha
\beta},q^{\dagger\dot\alpha \dot\beta}\}=0\ ,\ \ \{q_{\dot\alpha
\dot\beta},q^{\dagger\dot\rho
\dot\lambda}\}=2P^+\delta_{\dot\alpha}^{\
\dot\rho}\delta_{\dot\beta}^{\ \dot\lambda}\  ,%
\ee%
\bea\label{QQ}%
 \{Q_{\alpha\dot\beta},Q^{\dagger\rho
\dot\lambda}\}&=&2\ \delta_{\alpha}^{\ \rho} \delta_{\dot\beta}^{\
\dot\lambda}\ {\bf H} + \mu (i\sigma^{ij})_{\alpha}^{\ \rho}
\delta_{\dot\beta}^{\ \dot\lambda}\ {\bf J}^{ij} + \mu
(i\sigma^{ab})_{\dot\beta}^{\ \dot\lambda}\delta_{\alpha}^{\ \rho}
{\bf J}^{ab} \ , \cr
\{Q_{\alpha \dot\beta},Q^{\dagger\dot\rho \lambda}\}&=& 0 \ , \\
\{Q_{\dot\alpha \beta},Q^{\dagger\dot\rho \lambda}\}&=&2\
\delta_{\dot\alpha}^{\ \dot\rho} \delta_{\beta}^{\ \lambda}\ {\bf H}
-\mu (i\sigma^{ij})_{\dot\alpha}^{\ \dot\rho} \delta_{\beta}^{\
\lambda}\ {\bf J}^{ij} -\mu (i\sigma^{ab})_{\beta}^{\
\lambda}\delta_{\dot\alpha}^{\ \dot\rho} {\bf J}^{ab} \ .
\nonumber%
\eea%
The generators of the above supersymmetry algebra can be realized in
terms of $J\by J$ matrices as%
\bea P^+=-P_- = \frac{1}{R_-} \Tr {\bf 1} \qquad&,&\qquad
P^-=-P_+=-{\bf H} \cr q_{\alpha\beta}=\frac{1}{\sqrt{R_-}}\ \Tr
\theta_{\alpha\beta}\qquad&,&\qquad
q_{\dot\alpha\dot\beta}=\frac{1}{\sqrt{R_-}}\ \Tr
\theta_{\dot\alpha\dot\beta} \eea%
the rotation generators read%
\be\label{Jij}
 {\bf J}_{ij}= \frac{1}{2}\Tr
\big(X^{i}\Pi^{j}-X^{j}\Pi^{i}-2\theta^{\dagger\alpha\beta}
(i\sigma^{ij})_{\alpha}^{\rho}\theta_{\rho\beta} +
2\theta^{\dagger\dot{\alpha}\dot{\beta}}
(i\sigma^{ij})_{\dot{\alpha}}^{\dot{\rho}}\theta_{\dot{\rho}\dot{\beta}}\big)
\ee%
\be\label{Jab}
 {\bf J}_{ab}= \frac{1}{2}\Tr
\big(X^{a}\Pi^{b}-X^{b}\Pi^{a}-2\theta^{\dagger\alpha\beta}
(i\sigma^{ab})_{\beta}^{\rho}\theta_{\alpha\rho} +
2\theta^{\dagger\dot{\alpha}\dot{\beta}}
(i\sigma^{ab})_{\dot{\beta}}^{\dot{\rho}}\theta_{\dot{\alpha}\dot{\rho}}\big)
\ee%
%
\end{document}